# Rethinking Software Engineering in the Foundation Model Era

A Curated Catalogue of Challenges in the Development of Trustworthy FMware

Ahmed E. Hassan⋄, Dayi Lin⋆, Gopi Krishnan Rajbahadur⋆, Keheliya Gallaba⋆, Filipe R. Cogo⋆,
Boyuan Chen⋆, Haoxiang Zhang⋆, Kishanthan Thangarajah⋆, Gustavo A. Oliva⋆, Jiahuei (Justina)
Lin⋆, Wali Mohammad Abdullah⋆, Zhen Ming (Jack) Jiang•

cse@huawei.com

⋆ Centre for Software Excellence, Huawei Canada  ⋄ Queen's University, Canada  • York University, Canada

## ABSTRACT

Foundation models (FMs), such as Large Language Models (LLMs), have revolutionized software development by enabling new use cases and business models. We refer to software built using FMs as FMware. The unique properties of FMware (e.g., prompts, agents, and the need for orchestration), coupled with the intrinsic limitations of FMs (e.g., hallucination) lead to a completely new set of software engineering challenges. Based on our industrial experience, we identified ten key SE4FMware challenges that have caused enterprise FMware development to be unproductive, costly, and risky. In this paper, we discuss these challenges in detail and state the path for innovation that we envision. Next, we present FMArts, which is our long-term effort towards creating a cradle-to-grave platform for the engineering of trustworthy FMware. Finally, we (i) show how the unique properties of FMArts enabled us to design and develop a complex FMware for a large customer in a timely manner and (ii) discuss the lessons that we learned in doing so. We hope that the disclosure of the aforementioned challenges and our associated efforts to tackle them will not only raise awareness but also promote deeper and further discussions, knowledge sharing, and innovative solutions across the software engineering discipline.

## 1 INTRODUCTION

FMware refers to the type of software that uses foundation models (FMs), such as Large Language Models (LLMs), as one of its building blocks. Due to their general capabilities, flexibility, and natural language interface, FMs open doors to a new software development paradigm in which domain experts who do not possess deep programming or AI skills are empowered to build FMware for a variety of use cases [13]. The market size of FMware is estimated to grow at a compound annual growth rate (CAGR) of 35.9% from 2024 to 2030 [39].

Despite the benefits of FMs, building trustworthy FMware remains considerably difficult. As a global technology company with more than 200k employees and close to 100 billion US dollars in revenue in 2023, understanding the causes of this difficulty is essential to us. In our experience, we observe that developers constantly suffer from three key pain points when integrating FMs in software systems: (i) **low productivity** throughout the lifecycle, due to the lack of mature practices and integrated tooling, (ii) **high risk**, due to the stochastic and nondeterministic nature of FMs, and (iii) **high operational cost**, due to the high resource consumption of FMs.

And we are not alone. In a recent article by Microsoft [83], 26 professional software engineers responsible for building Copilot-like products echo our pain points: prompt engineering and testing are extremely time-consuming and resource-constrained. Additionally, the participants were frustrated with the lack of comprehensive tooling and best practices in this nascent domain.

In this paper, we introduce and discuss a curated catalogue of software engineering (SE) challenges for developing FMware (i.e., SE4FMware). These challenges were identified based on (i) surveys of academic and grey literature, (ii) in-depth discussions with industrial and academic leaders (e.g., during SEMLA 2023 [53] and the FM+SE Vision 2030 [43] event with over 100 attendees from many leading companies), (iii) meetings with our customers and our own development teams to understand the functional and non-functional needs of their FMware, and (iv) our practical experience in designing, implementing, and maintaining our in-house FMware lifecycle engineering platform (FMArts) and several complex FMware systems for strategic customers. For each challenge, we discuss the innovation path that we envision. These innovation paths call for significant breakthroughs in software engineering research and practice.

Subsequently, we introduce FMArts and its main components. FMArts was designed from the ground up to be a one-stop shop platform for the lifecyle engineering of trustworthy FMware. In particular, FMArts bootstraps productivity by introducing an opinionated development process that frees developers from having to choose between a plethora of individual tools and wasting time deciding how best to integrate them. Instead, FMArts lets developers focus on the real work that needs to be done (i.e., fulfilling functional and non-functional requirements).

This paper is structured as follows. Section 2 describes the novel aspects of FMware and motivates the need for innovation within the SE4FMware space. Section 3 discusses the software engineering challenges for FMware and associated innovation paths. Section 4 introduces FMArts and describes its main components. Section 5 describes the case study, including its context, the solution design, outcomes, and our learned lessons. Finally, Section 6 includes our final remarks.

## 2 SOFTWARE ENGINEERING FOR FMWARE

In this section, we first describe the software generation of FMware and how it relates to prior ones (Section 2.1). Then, we describe inherent limitations of FMware (Section 2.2). Finally, we go through our view of the FMware development lifecycle (Section 2.3).

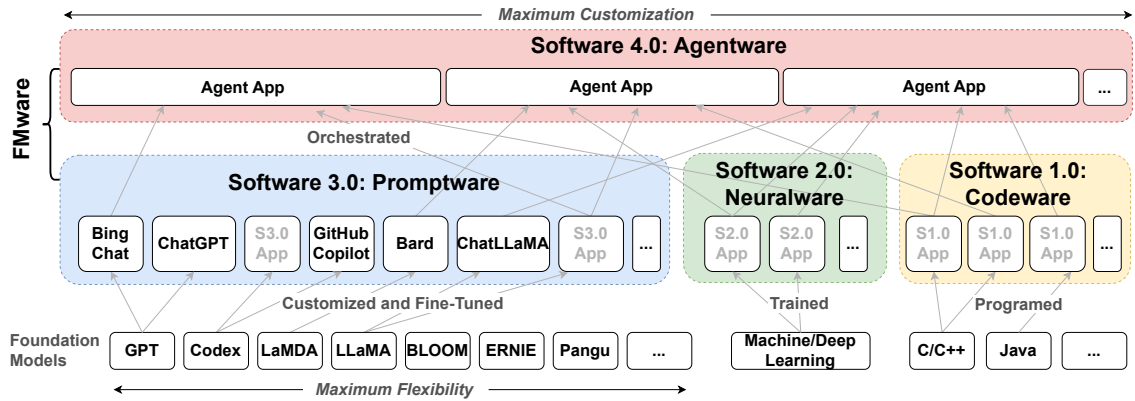

Figure 1: Agentware and its relation to prior software generations.

## 2.1 FMware in context

Over the past decade, we have witnessed the birth and evolution of several generations of software. Each new generation is characterized by drastic changes to the development paradigm, often involving new lifecycles, core assets, and roles. *Codeware* (a.k.a. Software 1.0) development involves professional programmers writing logic-based source code, following a typical development lifecycle that includes a well-established set of phases (requirements, design, implementation, testing, deployment, and maintenance) that are carried out either sequentially (e.g., the waterfall method) or iteratively (e.g., the agile family of development processes). As machine learning models started to be embedded in software in recent years, AI experts started to drive the development of a new generation of software known as *Neuralware* (a.k.a. Software 2.0) by constructing a new set of assets such as *datasets* and *models* in a significantly more iterative and experiment-driven lifecycle [6]. Since most of the traditional SE practices and tools were no longer applicable for Neuralware, many research efforts have been made in the SE community to address the unique challenges of Neuralware throughout its lifecycle (e.g., versioning of AI models [51]).

With the rapid development and adoption of foundation models, we are witnessing the emergence of yet another generation of software known as *Promptware* (a.k.a. Software 3.0). Different from the prior generations, Promptware is developed through *prompts* that are written in natural language. Such a paradigm shift unlocks the potential for software to be developed by *software makers*, i.e., people without deep programming and AI skills, ultimately democratizing software creation. ChatGPT is a popular example of Promptware.

Recently, people have started to leverage the reasoning capabilities of FMs to implement agents that are capable of decision-making, taking actions in an environment, and even interacting with other agents. In particular, agents achieve their goals by leveraging all prior software generations. We call this newest generation *Agentware* (a.k.a Software 4.0). AutoGPT [1], BabyAGI [2], CrewAI [3], and MiniAGI [4] are few examples of this recent development.

Finally, we foresee the birth of a new software generation once artificial general intelligence (AGI) becomes a reality. We refer to this generation as *Mindware* (Software 5.0). Mindware consists of a paradigm where development is led mostly by autonomous agents. Humans themselves are expected to intervene less in the software development. Instead, humans will focus on creating and refining *constitutions* [10] while overseeing the agents [59].

We collectively refer to Promptware and Agentware as *FMware* (due to the key role played by FMs in those software generations). Analogously, we collectively refer to Neuralware, Promptware, Agentware and Mindware as *AIware*. We also note that any given software generation typically includes software components from prior generations, and later generations of software will coexist with earlier generations rather than replace them. Figure 1 shows Agentware as an example.

## 2.2 FMware and the intrinsic limitations of FMs

Despite the huge market potential of FMware, FMs exhibit several *intrinsic limitations* that complicate the development of FMware:

• **Complex Task Limitations**: FMs are unable to handle complex multi-step tasks that require planning [98]. FMs also have limited reasoning capabilities [108] unless prompted in very specific ways (e.g., using prompt engineering techniques such as chain-of-thought prompting [101]). Finally, FMs exhibit uncontrollable behavior in multi-round interactions due to the limited size of the context window, and may get confused with long prompts even with growing context window size [63].

• **Hallucination Limitations**: FMs are prone to hallucination and can confidently generate factually wrong responses [48].

• **Closed Loop Limitations**: FMs by themselves are unable to interact with external resources [112].

## 2.3 The FMware development lifecycle

The distinguishing characteristics of FMware (Section 2.1) and the intrinsic limitations of FMs (Section 2.2) lead to a completely novel software development lifecycle. We summarize our view on the lifecycle of FMware in Figure 2. Such a view was consolidated based on our literature survey, meetings with our customers and development teams, and our practical experience in building both complex FMware as well as our own FMware lifecycle engineering platform. Yet, we acknowledge that such a view may not be complete due to the pace in which the field is evolving. Nevertheless, we hope that our concrete view will kick start urgently needed discussions around the FMware lifecycle, including the naming and scope of the various phases and activities.

**Requirements engineering.** The process starts at the *requirements engineering* phase (yellow box). In our experience, while requirements engineering should preserve some degree of technology agnosticism (i.e., focus more on the *what* than the *how*), we observe that a lack of understanding of the capabilities and drawbacks of fundamental technologies in the solution space (e.g., FMs) often leads to problems (e.g., unrealistic specification of system behavior). Hence, this phase involves thinking of requirements in the presence of FMs. For instance, FMware typically includes a strong conversational aspect and FMs lack deterministic behavior. Therefore, the elicitation and specification of requirements should account for these important characteristics.

**Cognitive architecture planning.** After requirements are collected, FMware goes into a novel phase that we call *cognitive architecture planning*. This phase entails choosing the coordination mechanism that will dictate what software systems (possibly from different generations) will be invoked, how, and in which order. In our mental model of cognitive architectures, the coordination mechanism choice is what draws the line between Promptware and Agentware. In Promptware, the coordination mechanism is a *static workflow* that prescribes a set of tasks and the order in which those will be executed (and a task can always be another workflow). These workflows are also known simply as *flows* (e.g., in Microsoft Promptflow [74]). We refer to the task of producing workflows as *workflow engineering* (see Promptware red box). While humans might leverage FMs to help them design (or co-design) these workflows, the end goal is still to obtain a static workflow. In the case of Agentware, the coordination mechanism is driven by autonomous agents. That is, agents decide the steps **and** the order in which those will be taken. In *multi-agent* cognitive architectures, coordination is achieved as a result of inter-agent interactions (e.g., Microsoft Autogen [105]). We refer to the task of designing agents (and their communication protocol in the case of multi-agent architectures) as *agentification* (see Agentware red box). In both Promptware and Agentware, task execution can always be accomplished by (or with the help of an) agent. It is also worth noting that although agents have higher autonomy, a certain level of human intervention is typically needed to reduce risks in FMware. For instance, *constitutions* [10] might be used to ensure that agents adhere to certain basic principles and ethical standards.

Architectural patterns [34] are starting to emerge for *cognitive architectures*. LangChain recently published a list of such patterns [58], which we adapt to our terminology and show in Table 1. The patterns are listed in order of controllability. For instance, in the first (topmost) pattern, the steps to be taken as well as their order are already established via a static workflow, thus giving humans quite a bit of control. In the last (bottommost) pattern, the steps to be taken as well as their order are dynamically determined by autonomous agents, giving humans little control (but more room for emergent behaviors and reasoning).

We believe that new cognitive architecture patterns will be identified over time and their trade-offs will become clearer as the field matures and different architectures start being deployed in enterprise settings.

**Table 1: Cognitive architecture patterns.**

| Type of FMware | Architectural Pattern | Decide Output of Step | Decide Which Steps to Take | Determine What Sequence of Steps Are Available |
|---|---|---|---|---|
| Promptware | FM Call | FM (one step only) | Workflow | Workflow |
| Promptware | FM Chain | FM (multiple steps) | Workflow | Workflow |
| Agentware | Router Agent | Agent | Agent | Workflow |
| Agentware | State Machine Agent | Agent | Agent | Workflow |
| Agentware | Autonomous Agent | Agent | Agent | Agent |

**FMware design, implementation & testing.** After the cognitive architecture is planned, the lifecycle jumps to a *design, implementation & testing* phase (blue box). This phase tends to be highly iterative (note the dotted back arrows). In our experience, developers typically jump between steps as the design evolves and requirements become clearer.

This xware *design, implementation & testing* phase starts with *interaction design* [85], which involves reasoning about and planning for the manner in which end-users will interface with the system. In the context of FMware, it is common that the input provided by end-users will be used as part of some prompt inside the application. In fact, many FMware systems prescribe a back-and-forth conversation between the end-users and the system (e.g., ChatGPT).

Next, we advance to the *(foundation) model engineering* task, which comprises selecting one or more FMs and aligning them if necessary. *Model alignment* is the process of modifying the parameters of an FM using a set of techniques and carefully crafted data, such that the resulting FMware is aligned with the requirements. Examples of alignment techniques include supervised fine-tuning [16] and reinforcement learning from human feedback (RLHF) [80].

Subsequently, we jump to the *cognitive architecture design* subphase. This subphase starts with the *prompt design* step. In this step, prompts are designed in accordance to the previously chosen cognitive architecture. Prompt design encompasses *prompt engineering*, *knowledge engineering*, and *context engineering*. Prompts are written using prompt engineering techniques that maximize the chances that the FM will consistently provide the desired response. Context engineering entails techniques for manipulating and enriching the context of prompts (e.g., prior interaction history and quick patches of FMware behavior). Knowledge engineering entails the creation, management, and retrieval of domain knowledge to ground the FMware. For example, grounding techniques such as retrieval-augmented generation (RAG) [60] have proven useful in mitigating FM hallucination. Once prompts are designed, the lifecycle moves to either the *workflow engineering* (in the case where coordination is dictated by static workflows) or the *agentification* step (in the case where coordination is dictated by agents). In our experience, prompt design and worflow engineering/agentification form a highly iterative loop, as one step drives and refines the other. As discussed above, developers might leverage cognitive architectural patterns to implement the chosen type of coordination. In complex FMware, the workflow engineering/agentification step typically involves invoking several software systems from different generations. For instance, an FMware could use a Python interpreter (Codeware) to execute python code, a neural network (Neuralware) to recognize patterns in images, an interactive copyeditor to support

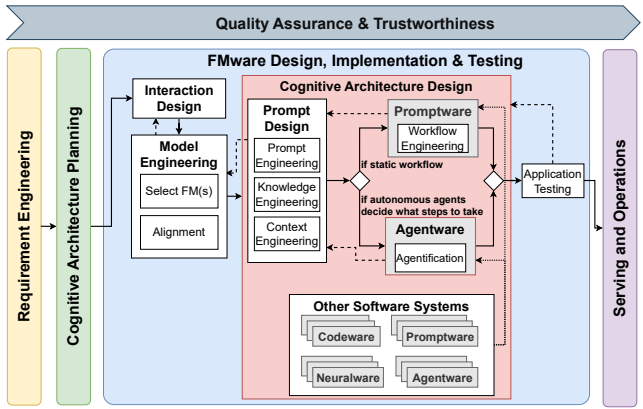

**Figure 2: Our view of the FMware development lifecycle.**

scientific book writing (Promptware), and a general task planner (Agentware).

Finally, the lifecycle advances to *application testing*. In this step, the whole FMware must be tested. Testing is particularly hard due to the non-determinism of FMs as well as the presence of heterogeneous multi-generational software (i.e., FMware is ultimately a complex *system of systems*).

**Serving and operations.** The last phase is *serving and operations* (purple box). This phase concerns the packaging, release, deployment, telemetry, monitoring, observation, and patching of FMware. Since the quality of FMware output is hard to quantitatively measure (see Challenge 9), traditional observability must be expanded to include the collection, processing, and analysis of data that represent end-user intent and experience, including explicit user rating of the FMware output, or implicit behavior such as correction of the output. We refer to this new type of observability as *semantic observability*.

**Quality assurance (QA) & trustworthiness.** Quality assurance (QA) & trustworthiness are cross-cutting and thus are enforced throughout the entire lifecycle (note the upper arrow in Figure 2). The intrinsic characteristics of FMs (e.g., non-determinism), as well as their limitations (Section 2.2), pose several challenges to QA and trustworthiness.

Overall, the unique characteristics of the FMware development lifecycle demand the research community to study and design SE principles and best practices for the development of trustworthy FMware. We refer to this new discipline as SE4FMware.

## 3 CHALLENGES IN ENGINEERING PRODUCTIONIZED FMWARE

In this section, we introduce ten SE4FMware challenges. As mentioned in Section 1, these challenges were identified based on (i) literature reviews, (ii) in-depth discussions with industrial and academic leaders, (iii) meetings with our customers to understand the requirements of their applications, and (iv) our practical experience in designing, implementing, and maintaining complex FMware systems for those customers.

To accomplish item (iv), our team created the FMArts platform. The FMArts platform is our long-term endeavor to design and implement a full-stack lifecycle engineering platform for the development of trustworthy FMware. As such, the challenges that we discuss in this section are also heavily influenced by our experience in developing our platform. FMArts is discussed in more details in Section 4.

Each challenge is tagged in blue with the FMware development lifecycle phase to which it belongs. Additionally, each challenge is tagged in red according to the pain point(s) that it causes, namely *low productivity*, *high risk*, and *high operational cost*.

### Challenge 1: Managing alignment data

⚙ Design, Dev. & Test | ❶ Productivity | ❶ Operational Cost

**Description.** To produce well-aligned models for trustworthy FMware, an alignment approach that seamlessly combines automation with strategic human input is essential. Pre-trained FMs often do not produce outputs aligned with the requirements and goals of enterprise use cases out-of-the-box. Therefore, aligning an FM with specialized enterprise data is essential to meet requirements and achieve market differentiation. Similar to Neuralware shifting development paradigm from code-driven to data-driven, aligning a model is achieved through data-oriented development, and needs to cope with challenges that are associated with data labelling, data debugging, data testing, and data versioning for FMware. Although Neuralware also requires human effort to manage training datasets, aligning FMs involves new types of data such as instructions, reference responses in natural language, ratings, and rankings of potential model outputs, requiring different efforts in the curation and maintenance processes. Another unique challenge for FMware is determining if a particular piece of data is part of the FM's training data and avoiding problems such as data leakage during evaluation. Consistent with recent industry reports [17], we observe that such processes demand significant time and financial resources. In addition, it is challenging to accurately and quantitatively gauge the adequate amount of high-quality alignment data for a specific use case, creating challenges for budgeting and time management.

**State of practice.** Researchers and practitioners are turning to active learning [96, 115] and weak supervision methods [9, 12] to reduce the involved effort in constructing alignment data. For example, Snorkel AI [89] employs data programming to generate labels using expert knowledge, while the Self-Instruct framework [100], a pioneering work in the area of auto-labeling, uses a powerful FM (e.g., GPT-4) to automatically generate labeled datasets from seed prompts, bypassing the manual labeling process.

Collecting alignment data in FMware is still too expensive and slow. Active learning methods demand significant manual input, with over 50% of the data needing manual labels [70], which is often unfeasible. Additionally, removing human labels can harm model alignment, as relying solely on automated labeling can lead to performance degradation, catastrophic forgetting, and model collapse [65, 94]. Another line of research attempts to reduce the needed amount of alignment data by focusing on the quality of the alignment data. Two pioneering works by Gunasekar et al. [41] and Zhou et al. [118] in this area show that fine-tuning an FM on textbook-quality alignment data results in an FM needing only a small amount of data for alignment. However, these methods still require a lot of manual effort to identify high-quality data. A recent effort by Zhang et al. [116] proposes an automatic method to identify a subset of high-quality alignment data within a larger dataset [41].

To support data debugging, researchers have employed statistical techniques (e.g., linear regression) to identify, select, and exclude undesired (e.g., biased) features, labels, and data points from training sets while ensuring the models' performance is on par (or better) than its counterpart trained in the original data [18, 61]. In the realm of FMs, influence functions and feature importance analysis can identify training data points that significantly impact the model's predictions [40, 92]. In addition, tools for data testing [14] either support the iterative process of cleaning up data while assessing its impact on models' performance [55] or automate such process [54]. However, such tools still need to account for the particularities of FMware, such as the lack of structure and the high dimensionality of the alignment data, including the opaqueness of FMs, which makes it difficult to determine the direct impact of training data on the model's outputs.

**Innovation path.** The management of alignment data for FMware calls for data IDEs that can fully support the model alignment lifecycle, including labeling, debugging, testing, and versioning data. Specific directions for such data IDEs include the integration of model- and human-in-the-loop, the programmatic labeling (data programming) and identification of inconsistencies or biases in the data, and features to support collaboration, such as supporting the aggregation and review of labeled data by different labelers. In our FMArts platform, we address some of these requirements by offering an *alignment data studio* that enables the curation of alignment data with an automated labeling approach that requires very little manual intervention. We also provide robust review mechanisms with support for debugging the labeled data. However, identifying biases, ensuring compliance, and testing the created alignment data remain a work in progress.

In addition, open and inner source data offer a significant window of opportunities to model alignment. Still, data may have licensing restrictions limiting their use, particularly in commercial applications. Therefore, such data IDEs should provide asset management capabilities for open and inner source data to reduce risks, for instance, by ensuring that the data is accurate, consistent, complete, and representative of a task. Also, as software makers and their domain expertise (Section 2.1) play a major role in data programming for FMware (e.g., by defining accurate labeling functions), data IDEs need to embrace the characteristics of its users, such as their potential lack of formal SE training.

## Challenge 2: Crafting effective prompts

Design, Dev. & Test | Productivity

**Description.** FMware needs an approach to prompt engineering that reduces human involvement and improves development efficiency and application quality, as well as prompt transferability and fragility. As the communication interface for FM, prompts at their current form are too low-level (i.e., at the level of input tokens to the FM) and fragile. This makes engineering prompts a non-trivial task due to the time-consuming and non-intuitive nature of iteratively hand-writing prompts [95]. This trial-and-error approach *lacks a clear structure* and *imposes a productivity bottleneck* on prompt engineering workflows, as multiple iterations of a prompt are created before the most suitable one for a given task is identified, particularly for complex functionalities that require chaining multiple prompts [106, 107]. To exacerbate the situation, *prompts are fragile*, in the sense that slight variations in a prompt can cause a significant change in the generated outputs [35], leading to continuous prompt adaptation and reevaluation, which might be costly if not properly managed. Similarly, small changes to the underlying FM or its hyper-parameters render existing prompts ineffective, ultimately hindering the generality of an implemented FMware solution. For example, as OpenAI constantly updates its served models [5], preliminary evidence shows that, when the same FM (e.g., GPT-4) is inferenced with the same prompt, the generated results by the FM differ at two different time instances [22].

Due to the unpredictable behavior of FMs, ensuring the accuracy and reliability of outputs becomes a significant challenge, as FMware developers must navigate the complex landscape of model behaviors without clear indicators of how undesired responses are being generated. Such lack of transparency and control makes achieving consistent and dependable results from FMware difficult.

**State of practice.** Visual tools [95] (akin to playgrounds) attempt to assist FMware developers by providing inference transparency (e.g., Prompt IDE [109] by xAI allows FMware developers to observe the generation process of FMs, such as sampling probabilities) and token-level explainability (e.g., Aleph Alpha [26] offers token-level attribution between input tokens and output tokens). While these capabilities are certainly helpful, they do not support higher level prompt debugging, i.e., identifying the logical units in a prompt lead to suboptimal results (e.g., a misleading few-shot example).

In addition, researchers and practitioners are actively looking for approaches to provide a cohesive and expressive framework for automating and supporting prompt engineering activities without forcing FMware developers to rely on completely manual, unstructured, or error-prone processes [64]. Several research endeavours attempt to fully [45] or semi-automate [46, 111] prompt engineering [114, 121] through *search-based approaches* or *reinforcement learning-based* approaches [27]. For example, Automatic Prompt Engineer [121] generates several instruction-based prompt candidates via a recursive process, and PromptBreeder [31] uses a genetic algorithm to select the most appropriate prompt. DSPy [52] combines a programming model with an optimization compiler to generate the best combination of prompts for a model. However, most approaches optimize prompts at the token level in isolation from other factors that might impact the results, such as model candidates, inference parameters, and hardware environments, while in our industrial setting, we see practices of FMware developers experimenting with a large set of FMs and fine-tuned variants, as well as prompt candidates under different settings to find the most suitable combinations.

**Innovation path.** To improve the quality of FMware output, researchers and practitioners have optimized the interaction with FMs in two directions: (i) breaking down one prompt for a complex task to prompting patterns that involve a series of prompts (i.e., prompt chains), sometimes with complex branching and conditions (e.g., chain-of-thought and tree-of-thought) and (ii) human involvement inside a chain of prompts to provide feedback to intermediate outputs (e.g., for a complicated code generation task, one could first prompt the FM to generate a description of the algorithm, then involve humans to correct the description, and prompt the FM again to implement the algorithm as code). Instead of focusing exclusively on prompt engineering techniques, researchers should also explore techniques for the design, debugging, visualization, and evolution of complex cognitive architectures that prescribe different levels of human involvement. Such techniques should incorporate higher-level prompts to support the dynamic selection and adjustment of lower-level prompts to sustain meaningful interactions based on the model's responses, context, and end-user interactions.

Designing complex cognitive architectures requires a proper *prompt IDE*. This IDE should support real-time collaborative features, such as sharing and managing prompt versions (including granular asset management support for the different parts of a prompt). These collaborative features are important to support teamwork and promote knowledge sharing. In addition, mechanisms to deal with just-in-time quality assurance should be provided, such as prompt debuggers and linters. In certain cognitive architectures, it is also important for prompt results to be reproducible. Yet, simply setting the FM temperature to zero is not enough in certain use cases. Hence, research must be carried out to create mechanisms that enforce full reproducibility [110]. Finally, to foster a rich and open ecosystem of functionalities, prompt IDEs should offer a plug-and-play architecture that facilitates integration with community-contributed tools and services, approaching the success of today's leading SE ecosystems like GitHub and Eclipse.

Towards accomplishing these requirements, we developed a *prompt studio* as part of our FMArts platform, providing an environment where version-controlled prompts can be connected to different FMs. This connectivity facilitates prompt refinement in conjunction with various FMs and execution setting configurations, including near real-time visualization of prompt optimization results. Our prompt studio offers asset management capabilities that enable prompts to be collaboratively developed, reviewed, and shared. Sharing can be either global (at the platform's scope) or local (at a project's scope, using role-based access control), thus facilitating teamwork. Our prompt studio also incorporates debugging capabilities, such as prompt interpretation tools that dissect the contributions of individual prompt components to the generated output. Token-level explanation is also supported for finer grain debugging. Finally, our FMArts also provides a *knowledge studio*. A key capability of our knowledge studio is to handle the upload, tokenization, vectorization, and retrieval of grounding data to be used in prompts, ultimately contributing to quality assurance (e.g., mitigation of model hallucination).

## Challenge 3: Multi-generational software

[Design, Dev. & Test] [Productivity]

**Description.** FMware developers face the intricate task of integrating different generations of legacy software systems, with each having its own data models, communication protocols and execution processes. This integration process demands supplemental effort to ensure seamless interoperability of many legacy software systems, thus hindering developer productivity. To maximize the usage of legacy software systems across multiple FMware and avoid redundant work, FMware developers need to either standardize the data formats of these legacy software systems or transform the data models at the post-retrieval step. However, such an integration approach introduces a significant workload. In addition, changes in the shared legacy software systems impact the downstream FMware, as it is sensitive to its input, i.e., prompts, as discussed in Challenge 2, increasing integration complexity. As such, FMware developers may opt to customize the legacy software for an individual FMware, ending up with maintenance overhead.

**State of practice.** To integrate multi-generational software, the current practices either use the concept of plugins to construct reusable units or package legacy systems as tools that FMware could reach. Semantic Kernel [73] leverages plugins through the construct of *native functions* to encapsulate the capabilities of a legacy system into functionality that can then be run by the kernel. Approaches such as LlamaIndex [68], ReAct [113], TaskMatrix [62], Toolformer [91], and ToolLLM [87] assist FMware developers to interact with legacy systems. However, the use of such techniques is often limited by the additional effort that is required from FMware developers to repackage existing tools or finetune FMs to be compatible.

**Innovation path.** Many current research efforts and frameworks assume FMware as a green-field project. Integrating with the large amounts of legacy software already in existence is not well addressed. The software engineering community should take into account how to effectively integrate FMware with legacy systems by considering *reusability* and *modularity* principles. In our FMArts platform, *cross-generational agents* are designed to reduce the effort of integrating multiple software generations in FMware, as well as to address evolution and compatibility concerns. This design objective is achieved by encapsulating behavior that is common to each of the software generations into different modules (which we call *skills*) and then unifying the interface of such modules into agents. Agents are designed using our *agent studio*.

## Challenge 4: Degree of controllability

[Design, Dev. & Test] [Serving & Ops.] [Risk]

**Description.** Trustworthy software systems, especially in sensitive domains such as finance, healthcare, and manufacturing, require robust controllability to prevent the possibility of misbehaviors causing unintended catastrophic consequences [102]. Researchers have emphasized the risk of AIware systems going out of control, even before the recent surge of FMs. The potential of systems that exert direct control over the world to cause harm in a way that humans cannot necessarily correct or oversee has been identified as a concrete problem in AI safety [7].

However, given the inherent stochastic nature of FMs (Section 2), determining the degree of controllability of FMware is challenging. It is critical to be mindful of the control of an FMware throughout its lifecycle. Specifically, we discuss controllability in three dimensions: (1) how, (2) when, and (3) who. FMware developers plan the control mechanisms of an FMware during the *cognitive architecture planning* and *cognitive architecture design* phases, both of which happen after requirements are elicited and analyzed. The *how* dimension includes determining the data granularity to protect sensitive data, access to various tools or services within FMware, and the autonomy to execute behaviors with or without human intervention. It is challenging to define the *how* aspect as FMware achieve certain tasks with different complexity. In terms of the *when* dimension, controllability of FMware varies across its lifecycle. FMware developers calibrate the degree of control of a given FMware over time and may disable certain control features if the FMware behaves unexpectedly in production. Lastly, the *who* dimension describes which stakeholders are authorized to grant control permissions, including FMware developers, end-users or other roles (e.g., admin). For example, during the execution of FMware, end-users grant permission to authorize an FMware to execute sub-tasks to achieve the high-level goals specified in the original plan.

**State of practice.** Restricted approaches, such as standard operating procedures (SOPs) and standard work instructions (SWIs), are employed for humans in industries where safety and quality are critical. Given the autonomous capability of FMware, there is an opportunity to make use of such existing approaches to effectively control FMware. Several frameworks have been proposed to customize the degree of control. Copilot Workspace [37] provides control by communicating the tasks that will be done in advance to the user so they can decide to continue on that path of action. AutoGen [105] is another framework that enables FMware development and offers mechanisms to solicit human inputs to offer different levels of control. Zhou et al. [119] propose to build controllable agents using symbolic planning. MetaGPT [44] leverages the assembly line paradigm and encodes SOPs into prompts to enhance human-encoded and FM-powered coordination, and thus controllability. All of these existing efforts and our own experience, highlight the importance of maintaining effective human oversight as a key requirement when supporting FMware.

**Innovation Path.** Fine-grained boundaries are defined in classic software systems that are designed to be extensible by third-party components (e.g., application permission model in Android [38] and the extensions architecture of Google Chrome [38]) when accessing resources to preserve the privacy and security of end-users. If elevated privileges are needed, components can explicitly request for permission from end-users. Similarly, FMware should leverage such a concept of boundaries to develop mechanisms to enforce constraints at a fine-grained level.

Existing concepts in software engineering, including *modularity* and *single responsibility principle*, are applicable to ensure FMware functionality is decomposed into small and independent components (e.g., agents) with well-defined interfaces. Therefore, component behaviours can be controlled in a granular manner. The need for abstractions to define control constraints facilitates the creation of consumable agents and tools for FMware developers. Our FMArts platform offers an *orchestration studio* for the creation of *SOP-based workflows* that constrain multi-agent behavior and interactions. Moreover, our platform auto-generates an architecture view by analyzing all assets in one FMware application. Developers can use this view to recognize deviations in implementation from the intended design. This holistic view aids in managing design principles such as *cohesion*, *coupling*, and *separation of concerns* within an FMware.

## Challenge 5: Compliance & regulations

[Requirements Eng.] [Design, Dev. & Test] [Serving & Ops.] [Risk]

**Description.** Trustworthy software systems are subject to stringent regulatory compliance requirements to ensure their security, safety, and compliance with relevant laws and regulations. Whilst, FMware is increasingly being used in decision-making processes where it is necessary to comply with such requirements. Despite governments all around the world having recently released regulations for FMs and FMware, it is still challenging to fully comply with them [42]. An example is the EU AI Act [30], which, in its Article 14, calls for high-risk FMware to be transparent and capable of being overseen by a human, despite this requirement being currently intractable for FMware.

A particularly relevant regulation is the GDPR in the European Union, which requires that personal and high-risk data should always be protected and should not leave secure bounds. For FMware that uses FMs as a third-party inference service, the possibility of data leaking beyond the secure enterprise bounds is high [83]. Several large enterprises like Microsoft and GitHub resort to self-hosting FMs to avoid such issues. However, not all enterprises have the means to self-host such large models. Therefore, SE4FMware needs to put forward measures, techniques, and tools to ensure regulatory compliance in such settings.

Ensuring licensing compliance within a FMware system is notably challenged by the opaque nature of FM development. FM providers do not always disclose their data sources [69, 93] or the usage of protected [19] and personal data [75], making it difficult for some FMware to comply with various data, code licenses, and copyright restrictions. Moreover, the necessity for diverse datasets and the concurrent operation of multiple FMs introduce significant complexities in license management, particularly when faced with ambiguous, conflicting terms [11, 88]. Notably, prominent FMware implementations like ChatGPT and GitHub Copilot have encountered legal challenges related to these compliance issues [24].

The high costs and extensive resources required to train FMs push enterprises towards utilizing open-source FMs. A prime example is LLaMA, an FM that was acclaimed for its open-source status when first released, yet restricting commercial usage and revoking licenses when exceeding 700 million subscribers[1] [72]. This issue is further compounded in the FMware development sphere by the increasing prevalence of non-Open Source Initiative (OSI) approved licenses for both FMs and datasets, undermining traditional open-source values and intensifying compliance dilemmas. Moreover, the extensive customization of these open-source FMs and datasets through fine-tuning and prompt engineering for enterprise-specific applications, including the development of progressively evolving agents, exacerbates the challenge of maintaining compliance amidst ambiguous open-source definitions.

---

[1] An updated version of the same model, called Lllama2, now allows unrestricted commercial use

**State of practice.** Efforts towards ensuring regulatory compliance of FMware remain largely manual and process-heavy. Brajovic et al. [15] propose to enhance the model and data cards that are currently used to document FMs with more relevant fields that could enable compliance with the EU AI Act. In addition, they propose two new schemas to document the relevant details of the FM's use case and deployment settings. However, they do not address how such reporting can be done for a complex FMware. As for data compliance, FMware can use a method that identifies copyright or license-protected data in FMs by analyzing the fingerprint of output tokens with those of protected data and FM outputs [93]. Another approach taken by several government bodies [78] and research initiatives [79] is to provide risk management frameworks that are applicable to FMware, which could, in turn, help ensure compliance with specific regulations.

License compliance efforts often concentrate on the individual FM or dataset, overlooking the intricacies of compliance within the FMware domain. A notable development in this area is the introduction of Responsible AI Licenses (RAIL) [25], aimed at incorporating behavioral and use-based restrictions into AI technology agreements to address ethical and regulatory challenges. Specifically, the RAIL-M licenses seek to embed AI-specific ethical standards through copyright licensing. Additionally, the Open Data Commons Public Domain Dedication and License (PDDL) and Creative Commons (CC) licenses strive to offer similar solutions for datasets [21]. Different from the licensing efforts, open source projects like OpenDataology [104] seek to decompose the dataset license into machine-readable AI-friendly use-case scenarios to enable better license interaction analysis. Despite these advancements, certain critical issues remain unaddressed, impeding comprehensive FMware compliance. These include focusing on singular models or datasets rather than integrated FMware systems, disregarding license interaction problems [21, 88], and overlooking the scenarios involving lifelong-evolving agents (e.g., learning a new behavior could violate that agent's current license).

Towards enabling FMware level compliance, recent initiatives to enhance Software Bill of Materials (BOMs) to represent AI and datasets offer a promising method for managing FMware and its dependencies (e.g., alignment data) in a format conducive to automated license compliance and vulnerability assessment. The advent of AIBOMs could facilitate the automatic risk analysis for FMware. For example, the release candidate for AI and Dataset BOM by SPDX [33], an ISO standard for SBOM, has a high potential for mitigating license compliance risks in FMware, as opposed to methods focusing on tracking the FM-related details only.

**Innovation path.** Tools capable of automatically determining FMware's composition and producing machine-readable SBOMs to catalogue FMs (FMBOMs) as well as every asset of an FMware (FMwareBOMs) are essential. It is also important to adopt well-established techniques such as testing and formal verification to analyze these FMwareBOMs and the FMware's output to ensure an FMware's regulatory and license compliance. Towards ensuring compliance, in FMArts, we implemented means to generate a preliminary version of FMwareBOM closely modeled after the SPDX 3.0's AI and Dataset profile [33]. With the help of these FMwareBOMs, FMArts conducts preliminary compliance analysis with enterprise rules and regulations using *automated reasoning* [90]. However, conducting formal verification on these FMwareBOMs remains an active area of development.

To mitigate the risk of legal consequences, redefining open-source for the FMware era and clarifying legal frameworks for agents is imperative. Moreover, the creation of FMware-specific licenses that consider the interplay between FM, dataset, code, and agent licenses is essential. Implementing tools akin to Fossology [103] tailored for FMware is pivotal in supporting enterprise FMware to remain compliant. Furthermore, documentation standards for disclosing compliance verification procedures and the shortcomings for FMware need to be established.

Another fundamental endeavor is the design of robust mechanisms to ensure enterprise and personal data privacy in FMware. As agents can acquire and exchange data with other agents, external services, or end-users, safeguard mechanisms should be systematically implemented to ensure that the incoming and outgoing data comply with regulations at all levels. In addition, despite telemetry information (e.g., logs) being of fundamental importance to understand FMware's behavior (including if the system is fully compliant), balancing the amount of information that is revealed to FMware operators is still an ongoing challenge that future developments in SE4FMware should tackle. In particular, privacy-preserving techniques can be applied so that operators get the most out of the telemetry data while preventing sensitive information disclosure, such as entire prompt details [83].

Finally, FMware has also been subject to new threats such as prompt injection attacks (e.g., jailbreak, toxicity) and adversarial attacks [66, 97, 120, 123], opening doors to malicious activities that can hinder security compliance. Even when guardrail protections are in place (e.g., via fine-tuning), FMware remains vulnerable to new threats that could be forged to bypass such protections. Therefore, the development of sophisticated tools and techniques to detect such threats is paramount to safeguard FMware applications.

## Challenge 6: Limited collaboration support

⚙ Design, Dev. & Test   ❗ Productivity

**Description.** For complex software, development is almost always collaborative. However, there is currently very limited support for the collaborative development of FMware. Such a limitation severely hampers team productivity, inner-source, open-source, and FMware ecosystem growth. Tackling this limitation involves rethinking the enablers of collaboration: from version control all the way to hubs (sharing platforms). In particular, the following points need to be addressed:

– *What to version and how.* In Codeware, we know that source code must be version controlled and the enabling technologies (e.g., Git) are very mature at this point. However, FMware introduces new types of assets. For instance, prompts are typically described using natural language and stored as raw text. As a consequence, logical units of a prompt (e.g., examples) across different prompts become a new type of reusable unit. In an effort to promote reuse, prompts are also sometimes written and stored as templates (e.g., using the Jinja2 template engine [86]). However, due to the prompt fragility problem (see Challenge 2), very similar prompts could yield very different results. Such a problem calls for version controlling all iterations where placeholders have been replaced. However, in that case, the connection between the template and its variants (as well as variants of variants) becomes lost, since hierarchical information is not natively preserved. More generally, the granularity at which prompts are represented, stored, and version-controlled must be carefully explored. Version controlling agents is also difficult due to the inherent characteristics of agents [99]. First, agents are often implemented as an amalgamation of source code, FMs, and potentially other assets. This brings up the question of how exactly agents should be serialized and version controlled. Moreover, intelligent agents are expected to incrementally acquire, update, accumulate, and exploit knowledge throughout their lifetime (a.k.a *continual learning* and *incremental learning*) [99]. In such a context where agents learn over time, as well as forget things that they had previously learned, it remains unclear *when* and *how* to version control the agent (e.g., when should snapshots be captured?).

– *What formats and standards to use.* The lack of standards and protocols for FMware assets significantly hamper interoperability, i.e., the ability to reuse assets created by different organizations. Additionally, the lacks of standards and protocols also limits the ability to design development tools that work with a wide range of assets (e.g., agents) out-of-the-box. As discussed by the AI Engineer Foundation [32], developers are building agents in their own way, making communication with those agents very hard since their interface is different every time.

– *How to package and distribute.* While significant advancements have been made in the packaging and distribution of AI models (e.g., via Hugging Face), it remains unclear what the stereotypical release of an agent or a prompt should look like. For instance, models in Hugging Face are versioned and include the datasets on which they were trained. Given the adaptive nature of agents, how should the "training data" (e.g., environment definition, actions on environment, signals received from environment, interactions with other agents and humans) be described and released? More generally, how does one package an agent? These release engineering problems need to be tackled to enable widespread sharing of assets and the future establishment of an ecosystem of FMware assets and creators.

**State of practice.** Prompts and prompt templates are typically recorded in Git. Simpler agents that encode a list of sequential steps are typically described using a markup language such as YAML (e.g., flows in Microsoft PromptFlow [74]). There are no established standards for describing prompts and agents. In the space of agents, one of the most prominent initiatives is the *agent protocol* proposed by the AI Engineer Foundation [32]. Such a protocol is being adopted by several relevant projects, such as AutoGPT [1] (a popular framework for building agents that focus on task automation). A committee within IEEE Standards Association (IEEE SA) is also currently working on an agent interface standard [8]. In terms of technology, while (i) Git has been established as the *de facto* technology behind collaborative development in Codeware and (ii) innovative technologies are being proposed to enable version controlling machine learning models in Neuralware [51], there is no established technology to enable collaborative development of FMware assets. Similarly, while established hubs exist for Codeware (e.g., GitHub) and Neuralware (e.g., Hugging Face), none exist for FMware assets.

**Innovation path.** Novel processes and technologies for versioning FMware assets need to be designed. Accomplishing this design requires determining what exact pieces of information need to be version-controlled and how frequently. In particular, the granularity at which prompts are defined must be examined carefully. Establishing standards and protocols for describing prompts and agents is a community endeavor that must be carried out to enable inter-organization reuse and interoperability. Once solutions to these problems are in place, a proper sharing platform (e.g., community-driven hubs) should be designed to foster the creation of a healthy ecosystem of high-quality FMware assets.

In our platform, we started approaching this challenge by enabling version control and granular role-based access control for all FMware assets. Particularly, in the case of prompts, each one of its logical component (e.g., instructions, examples, persona) is individually version- and access-controlled, ultimately enabling fine-grained asset management and reuse. In addition, all these components can be made available in our internal prompt hub, which can be accessed across an organization to enable inner-source initiatives.

## Challenge 7: Operation & semantic observability

⚙ Serving & Ops. | ❶ Productivity | ❶ Risk | ❶ Operational Cost

**Description.** FMware demands enhanced observability for a comprehensive understanding of its capability and behaviors since monitoring tools for classic software (e.g., tracking service status or execution logs) often fail to provide insights into the semantic outputs of FMware. Such enhanced observability is achieved by two approaches that we have coined as *enhanced operation telemetry* and *semantic signal telemetry*.

Enhanced operation telemetry includes monitoring for data drift, model parameters, token consumptions, groundedness, etc. during operations [82]. Such telemetry data enables analyzing the behaviors of FMware, identifying and ensuring the accuracy of the FMware, and preventing business-impacting issues when the data landscape changes, especially considering the inherently high cardinality of involved data. Furthermore, an FMware is typically inherently connected to multiple systems (e.g., legacy systems, databases, or even other FMware), and leverages the data produced by such a system as input to generate its own output, resulting in a complex system. Enhanced operation metrics in such a complex system help FMware developers navigate dynamic interactions among FMware and environments, and thus streamlines debugging activities. However, existing observability solutions for Codeware cannot provide comprehensive observability features for the unique metrics and complex interactions of FMware [29].

Semantic signal telemetry monitors explicit and implicit feedback of users to assess the quality of FMware in production, as the quality of FMware output is difficult to measure (see Challenge 9). Explicit user feedback often takes the form of a rating system (e.g., thumb-up/thumb-down or scale ratings). However, constantly asking end-users to provide explicit feedback to the FMware output is disruptive to the user experience. Implicit user feedback often involves monitoring end-users' reactions to the FMware output, such as the modification of generated results. However, defining accurate proxy behavior to user sentiment can be difficult.

**State of practice.** Recently many efforts have been made to enhance the observability of FMware. LangSmith, a closed-source platform from the popular FMware framework LangChain's community, provides observability in complex LangChain applications through logging runs, managing nested calls, and tracing component interactions [56]. However, achieving a level of observability that can comprehensively trace, monitor, and analyze the dynamics of FMware ecosystems, especially for Agentware with multiple agents beyond LangChain's single-agent and chain mechanism, remains a formidable challenge. Specific applications such as GitHub Copilot applies semantic signal telemetry by monitoring the editing behavior of users after a piece of code is generated, as an implicit feedback signal. However, there is no generalizable solutions for semantic signal telemetry for FMware.

**Innovation path.** Enhanced metrics of diverse aspects for observability are necessary to assist developers in understanding the capabilities, analyze behaviors, and ensure the performance of FMware, given the dynamics and complexity of FMware. Our FMArts platform includes an *operations studio* that facilitates observability by supporting the monitoring, logging, and alerting of metrics that are unique to FMware. However, more efforts need to be made to advance generalizable semantic signal telemetry.

## Challenge 8: Performance engineering

[Serving & Ops.] [Productivity] [Operational Cost]

**Description.** Deploying and serving enterprise FMware introduces significant performance challenges, mainly due to the vast size of the underlying FMs and the need for multiple, high-performance interactions [13]. In line with recent research, we find that larger prompts or more complex FMs increase inference time and performance drops with longer prompts from multi-round conversations, which are common in enterprise FMware. Furthermore, when FMware uses externally-hosted FMs, larger prompts or multi-round conversations involve increased token usage and, as a consequence, the cost and latency of operating the FMware [49] increases. This is problematic due to token limits/context length and inference speed throttling enforced by FM providers, which impact the performance of the deployed FMware. Thus, tackling these performance issues is crucial for successfully implementing trustworthy FMware.

In addition, in an enterprise context, an FMware may involve multiple FMs and multiple rounds of inference for each FM. The number of rounds of inference may not be predictable before execution due to the potential dynamic nature of FM-based planning and orchestration. Hence, even if the performance of each FM is optimized and meets the model level Service Level Agreements (SLAs), the FMware may still fail the application level SLA. The problem compounds in an Agentware, where each agent is expected to continuously plan, learn, and evolve, requiring unforeseeable amounts of inference from multiple FMs.

Another key concern in an enterprise setting is that when the FMs are self-hosted on an enterprise's private cloud, the computing resources (e.g., GPUs) available to a given FMware are often limited. In such a case not all needed FMs can be loaded into GPUs at the same time. Loading and off-loading FMs by demand are therefore needed and could take tens of seconds each time, posing significant performance challenges to FMware. Furthermore, FMware, which is composed of multiple agents, needs a data dependency-oriented execution plan since one agent might depend on the outcome of another agent. In such a setting, optimally loading and unloading the FMs required by different agents into the GPU, given that some FMs may have a cold start problem for a given agent (since the FM may not be finetuned) within the constraints of the FMware's SLA, poses significant performance challenges to enterprise FMware.

**State of practice.** Much of the existing research focuses on solving performance bottlenecks at a single-round inference level (i.e., FMOps level). There is little progress made in the FMware level with multiple models and multiple rounds of inference (i.e., FMwareOps level). One strategy used to ameliorate performance issues in FMware level is selecting FMs with fewer parameters to power the FMware. For instance, in Meta, when building an FMware for code authoring, they purposefully choose the 1.3B InCoder FM over the 6.7B model to avoid inference latency despite the drop in accuracy [76]. Another method involves compressing the FMs through quantization, distillation, or pruning, reducing the FM's size to improve inference time [122]. However, how to effectively compress the FM without sacrificing its performance remains a challenge. Additionally, researchers explore prompt compression [50], aiming to shorten input prompts while preserving essential information, though such practice risks losing critical semantic details, especially in logical reasoning tasks. A promising area is semantic caching [36], which stores responses to previous prompts for quick retrieval in similar or identical queries, enhancing the efficiency and speed of enterprise FMware interactions.

**Innovation path.** Existing practices treat FM inference acceleration as a standalone endeavour and FMware is parsed into a series of imperative statements used in common programming languages (e.g., Python). Such a representation will lose the intent behind developers' reasoning about optimal task placement associated with data locality, leaving the optimization made only by compilers. To cope with this challenge, we propose that FMware should be defined and presented in a declarative manner with a higher level of abstraction. Such abstractions could capture developers' intent and expose advanced application graph-level optimization [117]. In addition, optimizing the performance at FMware level also takes into account FM characteristics, hardware heterogeneity, and data movement cost, providing a global performance view compared with accelerations of individual FM inference speed. More research should be conducted towards these joint optimizations between the front-end intent-aware representation, the back-end application graph performance optimization, and the FM inference accelerations. Our FMArts platform represents the developed FMware in a declarative manner. In FMArts, the SOP-based workflow (either created manually or by an FM) specifies the operating procedure, which is represented as an execution graph. This execution graph is passed on to a runtime component called *fusion runtime*, which determines the optimal way to execute the SOP-based workflow through various graph and performance optimizations (e.g., elastic resource auto-scaling). These optimizations are geared towards decreasing the FM inference latency and data movement, as well as maximizing hardware utilization while still meeting SLAs. Ultimately, our fusion runtime increases the overall performance of FMware systems by efficiently using the hardware. Finally, the fusion runtime also supports a special type of cluster, which we call *uni-cluster*. A uni-cluster is a single cluster that is specially designed to serve multiple purposes, such as FM training, fine-tuning, serving, and FM-based agent customization. In our experience, uni-clusters drastically simplify the management of computing resources, resulting in increased operations (Ops) team productivity.

## Challenge 9: Testing under non-determinism

[Design, Dev. & Test] [Productivity] [Risk]

**Description.** Unlike the conventional AI models in Neuralware which are often used in predictive tasks (e.g., classification, regression), FMware often utilizes FMs in tasks that are generative in nature (e.g., content creation, reasoning and planning). Such difference poses two unique challenges in evaluating the quality of FMware. First, there exists no ground truth for a given input. Predictive tasks often have a sole ground truth that can be compared against and allow quantitative evaluation of the quality of the Neuralware. On the contrary, generative tasks often have more than one "correct" answer, which can be semantically similar (e.g., for translation tasks) or drastically different (e.g., for creative writing tasks). Hence, it is challenging to define clear "pass/fail" criteria for FMware. Second, now every test needs to deal with non-determinism and becomes flaky [83, 84]. Ensuring the reproducibility of FMware is even more challenging than that of Neuralware, invalidating the premise of using test suites to evaluate the quality of FMware as it is neither practical to list all potential variants of acceptable outputs [77], nor will the testing results remain consistent across different runs. In addition, testing FMware incurs substantial costs, as every request to an FM carries a financial burden (e.g., 1-2 cents to run [83]).

**State of practice.** The current practices of evaluating the quality of outputs from FMware either rely on human efforts, or involve automated validation against a reference solution using another high-performing FM such as GPT-4 [67]. For example, LangChain documentation suggests using several FM-based evaluators and ensuring reproducibility by setting a low temperature [57]. But even setting the temperature to zero does not guarantee determinism of the evaluator output [81]. Moreover, leveraging GPT-4 as an evaluator to assess the quality of outputs generated by FMware only achieves a correlation of 0.51 with human evaluation result [67], posing risks on the reliability of the evaluation result.

**Innovation path.** Novel approaches to define and evaluate the quality of FMware, with reduced human effort yet high reliability are needed. FMArts offers experiment and testing features to evaluate multiple FMware variants (e.g., prompt, model, and inference parameter combination) against a testing set of inputs. However, it is still challenging to quantify and summarize the test results without heavy human evaluation efforts.

## Challenge 10: Siloed tooling & lack of process

[All phases (cross-cutting)] [Productivity] [Risk] [Operational Cost]

**Description.** The complexity of developing FMware has led to an explosion of siloed tools that only tackle a specific aspect of the lifecycle [83]. For instance, there exist individual tools to support data alignment (e.g., Snorkel AI [89]), prompt engineering (e.g., xAI PromptIDE [109] and IBM PromptIDE [95]), and agent orchestration (e.g., Microsoft AutoGen Studio [105]).

This siloed tool setup slows down development due to cognitive overload, context switching, and the need to implement integration (glue) code for several interfaces that are managed by different offering providers. Consequently, the pace at which software solutions can produce business value is also slowed down. In our experience, the siloed tool setup also hinders the trustworthiness of FMware applications, as the lack of standardized development tools and practices lead to non-automated and inefficient compliance & governance.

**State of practice.** No lifecycle engineering solution currently exists for developing FMware [83]. We do acknowledge, however, the ongoing efforts of key industry players (e.g., PromptFlow by Microsoft [74] and LangSmith [47] by Langchain). In the research space, we also observe a few initiatives, such as (Bee)* [71] and Prompt Sapper [23]. However, current solutions either cannot cover the full lifecycle of FMware, missing critical considerations such as compliance and trustworthiness, or offer little extensibility and flexibility to integrate richer functionalities from the community.

**Innovation path.** Scalable and efficient development of FMware can only be accomplished by means of a truly unified platform that provides cradle-to-grave lifecycle support and offers flexibility for choosing between capabilities and integrating them. Our platform FMArts represents our effort towards achieving this goal.

## 4 FMARTS: OUR FMWARE LIFECYCLE ENGINEERING PLATFORM

Since we wish to create a stable foundation upon which all features and services are developed, we started the design and implementation of FMArts from scratch using a bottom-up approach. An overview of FMArts is shown in Figure 3.

In the following, we describe the layers of FMArts in detail, namely the FMware framework (Section 4.1), the FMware hub (Section 4.2), and the FMware graph compiler & fusion runtime (Section 4.3).

### 4.1 Framework

Our framework encompasses a set of *assets*, which are the primitive building blocks in FMArts. Assets can be *cross-generational agents*, *structured prompts*, or *SOP-based workflows*.

**Cross-generational agents** are designed to reduce the effort of integrating multiple software generations in FMware, as discussed in Challenge 3. In FMArts, an agent has the following components:
- *SkillSet* describes a set of skills that an agent has, with each skill containing a customizable implementation of an agent's functionality. An OpenAPI specification describes how to access each skill contained by the agent. Four different types of skills can be associated with an agent. *NativeSkill* focuses on Codeware and offers abstractions to either interpret native code (e.g. Python) or access an API endpoint. *ModelSkill* focuses on Neuralware and offers abstractions to perform inference on a conventional ML model using tabular data as input. *SemanticSkill* focuses on FMware and offers abstractions to perform inference on an FM using prompts as input.
– *AgentCard* contains a natural language description of an agent's metadata (e.g., licensing and known risks) and functionalities, being used to documentation and to guide FM-based orchestrators or humans to decide which skill should be used in a workflow that implements a particular behavior.

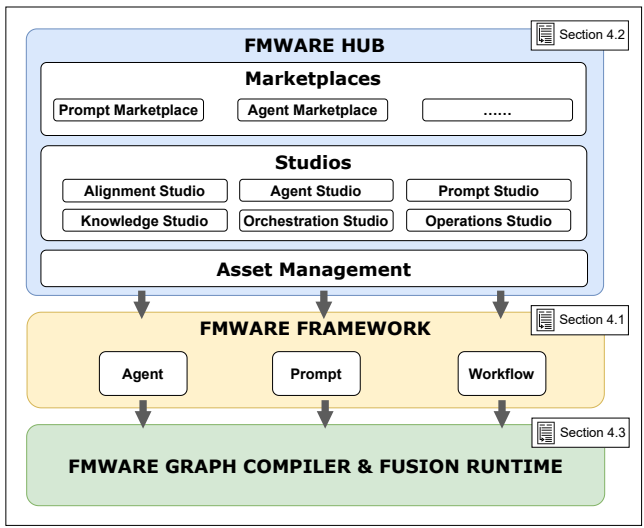

Figure 3: An overview of FMArts.

In FMArts, agents are designed using our *agent studio* (Section 4.2.2).

**SOP-based workflows** enable the completion of tasks using the skills of one or more agents, specifically tailored to tackle associated challenges in controllability (Challenge 4). Workflows incorporate familiar constructs such as sequencing, branching, and loops that can be created manually or with the assistance of FMs, and then automatically checked to ensure adherence to SOP descriptions. One workflow asset can reference multiple other workflows, promoting reusability and modularity.

**Structured prompts** are designed to tackle the unstructured nature of prompt engineering (Challenge 2), emphasizing modularity and reusability of different prompt components (e.g., personas, examples, and instructions) [23] by allowing each component to be maintained in separation from others. This design principle promotes a structured development approach, where individual prompt components can be optimized, refined, updated, managed, and reused independently.

### 4.2 FMware Hub

This component encompasses tools and infrastructures to support the FMware engineering lifecycle described in Section 2. FMware Hub is organized around marketplaces and studios, both relying on the asset management service.

*4.2.1 Marketplaces.* Marketplaces provide a means for prompts, agents, and other FMware assets to be shared and reviewed. The access scope of a marketplace can be at the team level, department level, organization level, or fully public.

*4.2.2 Studios.* To mitigate the siloed tooling problem (Challenge 10), FMArts offers a collection of low-code user interfaces with a homogeneous user experience for developing, maintaining, and integrating FMware assets. There are five such studios currently implemented in FMArts (Figure 3):

**Agent Studio** allows developers to define skills, associate one or more skills to an agent, and describe an agent's metadata.

**Prompt Studio** represents our ongoing efforts towards creating an environment for crafting effective prompts (Challenge 2). Key features include asset management at the prompt component level (e.g., persona, context, and instructions), visualization of prompt optimization results, prompt debugging, and support for collaborative development.

**Alignment Studio** is our partial response to the challenges related to data alignment (Challenge 1). The FMArts Alignment Studio is an environment for curating alignment data that employs varying degrees of automation (fully automated or with human-in-the-loop) and enterprise-grade compliance measures.

**Knowledge Studio** allows seamless creation, indexing and storing (e.g., tokenization and vectorization for vector-based knowledge management), and retrieval of grounding data (e.g., textual documents with up-to-date, domain-specific knowledge) and the integration of such data with other platform assets (e.g., loading retrieved contextual data into a prompt) when developing knowledge-intensive applications. Knowledge management has an important role in supporting effective prompting (Challenge 2), as it facilitates the implementation of grounding techniques.

**Orchestration Studio** allows developers to compose agents' skills into workflows and review the orchestration plan provided by an FM, enabling us to mitigate Challenge 4. In addition, our orchestration studio allows one to automatically deploy such workflows as web services.

**Operations (Ops) Studio** provides a semantic observability component to facilitate monitoring of reliability and performance aspects of FMware. Upon entering and exiting a workflow's node, the system generates detailed logs containing timestamp information, exit status, or stack traces in case of errors during execution. These logs also encompass records of external APIs access, resource consumption metrics (such as memory usage, CPU utilization, and latency), and the number of tokens generated. Additionally, the logging mechanism captures parameter values, including model configuration parameters, offering a holistic view of the system's behavior throughout its execution. Developers can set thresholds for specific metrics that, when reached during a node's execution, trigger notifications.

In FMArts, quality assurance (QA) is directly built into the Studios (4.2.2) instead of being treated as an afterthought. This design choice ensures that quality concerns are dealt with as soon as possible and do not ripple through the system. For example, the FMArts Orchestration Studio has a built-in compliance verification mechanism that checks whether each node complies with a set of pre-defined rules (e.g. whether all outgoing data is in the public domain or incoming data is authorized).

*4.2.3 Asset Management.* To solve some of the challenges associated with the lack of collaboration support (Challenge 6), FMArts Hubs uses a combination of a central repository with local workspaces to record version-controlled assets, enabling collaborative development among developers engaged in the same project using a "push-edit-and-pull" model.

### 4.3 Graph Compiler & Fusion Runtime

Importance of SLA becomes paramount when FMware is deployed at scale. For instance, Meta [28] mentions many optimizations in their serving approach that are geared towards meeting SLAs. The goal of our *graph compiler & fusion runtime* is to provide many such optimizations in a transparent and reusable manner to improve developer productivity as well as ensure the performance portability and evolution of FMware (as the FMware itself changes or the underlying associated hardware).

To this end, FMArts starts the execution of the FMware by compiling the workflow and its constituents into a graph representation. The compilation process preserves the original control flow in the generated FMware graph. Unlike other frameworks where the FMware is expressed using *inductive* expressions (e.g., .py files), the graph representation allows for a *declarative* expression of the FMware logic and preserves rich intent information that allows our execution environment (*fusion runtime*) to efficiently carry out parallel executions and dynamically allocate resources to optimize execution time and hardware utilization. Furthermore, fusion's multi-purpose clusters (uni-clusters) facilitate resource management and thus improve the productivity of operations teams.

## 5 CASE STUDY

This section describes one case study that we implemented purely using FMArts as per a customer request. The implementation discerns the capabilities of FMArts that are described in Section 4 on the path to tackle challenges presented in Section 3.

### 5.1 Context

**The customer.** Company A (anonymized due to confidentiality) is a large-scale e-commerce vendor that operates digital storefronts across multiple countries. The company enables customers to seamlessly browse product catalogs, purchase goods, and track deliveries via web and mobile applications. The company boasts a net annual revenue of over 50 million US dollars and has more than 15,000 employees. Moreover, Company A offers a portal where customers can engage with the support team , dealing with issues such as product specifications, offers, existing orders, and shipping details.

**Requirements.** Company A needed to build an *Intelligent Customer Support (CS) Specialist* that is capable of addressing user inquiries and escalating issues to the human CS team when needed. The intelligent CS specialist needs to follow the predetermined SOPs that human CS team follows and has the capability of interacting with several legacy systems in the company. The predetermined SOPs consisted of over 39 intents (e.g., modifying an order, invoice issues) for user inquiries and the corresponding procedures for each of the intents. The goal was to reduce the workload of the human CS team and reduce operational costs of the company while leveraging the company's rich internal knowledge base. The databases included detailed guidelines in natural language for handling various types of user requests, up-to-date product information, and the order details of a given customer. There are APIs available to retrieve such data from internal DBs or to programmatically trigger actions to a particular legacy system, depending on user requests (e.g., cancel an existing order or place a new order).

### 5.2 Solution Design

After understanding FMArts' features and analyzing the requirements of Company A, a team of three developers, with at least two years of experience in software engineering and no experience related to FMware, started to implement a prototype of the intelligent CS specialist using FMArts. The developers collaboratively created the required assets, such as prompts, agents, and workflows using the relevant web-based studios (Section 4.2.2) provided by FMArts. The team iteratively experimented and enhanced individual assets in two days, then integrated the assets into a workflow in another three days (*i.e., a total of five days*). Subsequently, after testing the end-to-end solution, the final FMware graph was deployed into the FMArts Fusion runtime (Section 4.3), exposing the application as a RESTful web service.

Figure 4 shows the architecture diagram of the implemented assets, which includes four agents, three prompts, three (web service) APIs, two FMs, and the encompassing workflow. The multitude of components on the left side in Figure 4 demonstrate how FMArts is able to integrate multiple generations of software when executing a task (Challenge 3). The steps illustrated in the workflow on the right-hand side of Figure 4 closely follow the SOPs that are used by human CS staff to respond to queries, addressing controllability (Challenge 4) and corporate compliance (Challenge 5) requirements.

### 5.3 Critical Features of FMArts

Company A was satisfied with the outcome delivered in such a short time span (i.e., within five days). The following unique features of FMArts were critical to achieve a successful delivery of the FMware:

**Mitigating hallucinations and non-deterministic behavior.** By adding a validation step after the intent identification step in the workflow, as well as post-conditions in the prompts, we avoided erroneous FM outputs (e.g., outputs that violate the predefined intents). Such capability was enabled by the Prompt and Orchestration Studios (Section 4.2.2).

**Flexibility to choose FMs.** Different FMs perform better in different types of tasks and have different operational costs. In the case study, by experimentation, the developers identified that the Alpaca model works better for the translation task and that LLaMA2 performs better for the intent classification task. Therefore, they opted to use these two FMs in the construction of the intelligent CS specialist, as supported by FMArts through its hubs (Section 4.2.3).

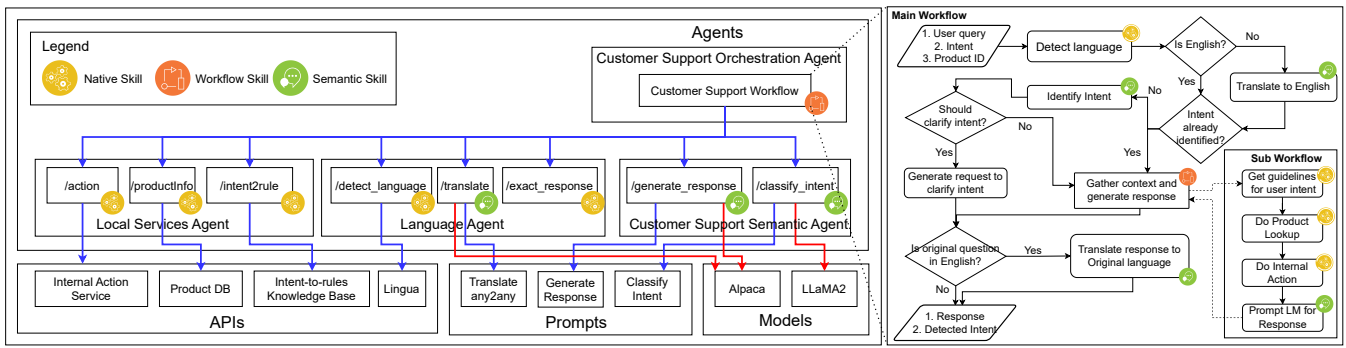

**Figure 4: Architecture diagram of the intelligent CS specialist which demonstrates `FMArts`' capability to build FMware with heterogeneous assets and workflows supporting SOPs. *Note: `FMArts` can autogenerate architecture diagrams and track them. This figure was redrawn by hand for clarity and to remove sensitive information.**

This observation highlights the importance of facilitating experimentation with multiple FMs and offering flexibility for choosing the suitable FM to match the requirements of an FMware.

**FMware development as a collaborative endeavor.** Each developer has their own skill sets and specialized areas. Therefore, development tools should provide features for a team of developers to collaborate on one project seamlessly. In the case study, the three developers worked on constructing different components (e.g., one worked on the agents while another worked on testing appropriate prompts) of the intelligent CS specialist in parallel. This was facilitated by the collaboration features and role-based access control in the `FMArts` platform (Section 4.2.3)

### 5.4 Lessons Learned

In the following, we discuss the lessons that we learned from having carried out the case study. These lessons helped us identify the current limitations of `FMArts` and the need for further research:

**Debugging difficulties at runtime.** The development of the intelligent CS specialist involved coordination among multiple FMware components, resulting in two groups of debugging challenges: 1) validation of input and output data for each of the `FMArts` components to ensure that the FMs behave as expected, and 2) identifying performance bottlenecks and optimizations such as reducing overall execution time (e.g., by caching intermediate steps). To tackle the first group of challenges, we identified the need for improving observability (Challenge 7) by adding support for collecting and visualizing relevant runtime metrics during execution. This feature provides granular insights into the behavior of individual `FMArts` components and facilitates the identification of logical errors and other issues (Challenge 9). Addressing the second group of challenges involves analyzing past execution data, identifying opportunities to optimize, and providing developer hints to refactor FMware (Challenge 8). We believe investing resources into researching these aspects further will help to increase developer productivity in the long run.

**Testing prompt variants exhaustively.** In the case study, the FMs are used not only for generating responses to user inquiries but also for tasks such as translating inquiries between languages. The diversity of these tasks highlights the need to overcome the inherent limitations of FMs. For instance, the customer observed instances where initial versions of the intelligent CS specialist struggled to accurately categorize the user inquiry into one of the predefined intents or escalate issues due to contextual shifts in translation. To address this issue, features of the Prompt Studio (Section 4.2.2) were invaluable, such as its ability to experiment with prompts and mitigate hallucinations by filtering FM outputs. However, further research into end-to-end testing [20] of FMware in the presence of many moving pieces will provide value in enterprise scenarios (Challenge 9).

**Performance optimization.** As FM inference is costly (Challenge 7), we opted for a more economical alternative by employing an FM with a reduced set of parameters (e.g., llama-2-7b) during development time and mirroring the characteristics of the model in the production environment (e.g., llama-2-70b). Despite the similarities, subtle variations in behavior were observed between the two models, even when prompted with the same input. This demonstrates the difficulty of how to cost-effectively develop an FMware and assure the outputs of the FMware in production (Challenge 7 and Challenge 9). Research into prompt portability between small and large models [41] will be helpful in this regard.

We expect these lessons to help pave the way for further research in these areas by the software engineering community.

## 6 CONCLUSION

The goal of this paper is to raise awareness of SE4FMware challenges that are observed in an industrial setting. But where should we go from here? Fully tackling those challenges requires the design of breakthrough technologies, techniques, and standards. We hope that our ten challenges will foster further discussions and knowledge sharing between software engineering researchers in industry and academia. In particular, we believe that certain challenges such as *limited collaboration support* (Challenge 6) can only be overcome by creating community-wide initiatives (e.g., a strategic consortium and interest groups) that carry out open discussions and prescribe the involvement of the broader community of software developers, industries, and open-source. More generally, we hope that the disclosure of our challenges, progress, and solutions will foster further explorations and discussions around the many challenges of this nascent and yet important domain of FMware.

## ACKNOWLEDGMENTS

The findings and opinions expressed in this paper are those of the authors and do not necessarily represent or reflect those of Huawei and/or its subsidiaries and affiliates. Moreover, our results do not in any way reflect the quality of Huawei's products.

We would like to acknowledge the following people's contributions to the development of FMArts (ordered by last name): Jack Basha, Aaditya Bhatia, Charles Chang, Ximing Dong, Azmain Kabir, Hao Li, Yu Shi, Cedric Wang, Shaowei Wang, Xu Yang, and Sky Zhang.